\documentclass[twocolumn,aps,pre,showpacs,floats,notitlepage]{revtex4-1}
\usepackage[usenames,dvipsnames]{color}
\usepackage{amsmath}
\usepackage{amssymb}
\usepackage{graphicx}
\usepackage{anysize}
\usepackage{mathrsfs}
\usepackage{notes2bib}
\usepackage{multirow}
\usepackage{mathtools}

\marginsize{1.5cm}{1.5cm}{0.5cm}{1.cm}
\usepackage[unicode=true,pdfusetitle, bookmarks=true,bookmarksnumbered=false,bookmarksopen=false, breaklinks=false,pdfborder={0 0 1},backref=false,colorlinks=true]{hyperref}
\hypersetup{colorlinks=false,
	linkcolor=red,
	anchorcolor=brown,
	citecolor=blue,
	urlcolor=black
}

\usepackage{breakurl}

\allowdisplaybreaks
\makeatletter
\@ifundefined{textcolor}{}
{
 \definecolor{BLACK}{gray}{0}
 \definecolor{WHITE}{gray}{1}
 \definecolor{RED}{rgb}{1,0,0}
 \definecolor{GREEN}{rgb}{0,1,0}
 \definecolor{BLUE}{rgb}{0,0,1}
 \definecolor{CYAN}{cmyk}{1,0,0,0}
 \definecolor{MAGENTA}{cmyk}{0,1,0,0}
 \definecolor{YELLOW}{cmyk}{0,0,1,0}
}

\usepackage{graphics}
\usepackage{epsfig}
\usepackage{epsf}

\providecommand{\nn}{\nonumber}
\providecommand{\be}{\begin{equation}}
\providecommand{\ee}{\end{equation}}
\providecommand{\bea}{\begin{eqnarray}}
\providecommand{\eea}{\end{eqnarray}}
\providecommand{\beas}{\begin{eqnarray*}}
\providecommand{\eeas}{\end{eqnarray*}}

\providecommand{\beni}{\begin{equation*}}
\providecommand{\eeni}{\end{equation*}}

\providecommand{\bw}{\begin{widetext}}
\providecommand{\ew}{\end{widetext}}
\providecommand{\imi}{\mathrm{i}}
\newcommand\EatDot[1]{}

\makeatletter
\newcommand{\vast}{\bBigg@{3}}
\newcommand{\Vast}{\bBigg@{4}}
\makeatother

\makeatother

\begin{document}

\title{Non-perturbative effects in spin glasses}

\author{Michele Castellana}
\affiliation{Joseph Henry Laboratories of Physics and Lewis--Sigler Institute for Integrative Genomics, Princeton University, Princeton, New Jersey 08544, United States}
\author{Giorgio Parisi}
\affiliation{Dipartimento di Fisica, Sapienza Universit\`{a} di Roma, Piazzale Aldo Moro 5, 00185, Rome, Italy}


\begin{abstract}
We present a numerical study of an Ising spin glass with hierarchical interactions---the hierarchical Edwards-Anderson model with an external magnetic field (HEA). We study the model with Monte Carlo (MC) simulations in the mean-field (MF) and non-mean-field (NMF) regions corresponding to $d\geq4$ and $d<4$ for the $d$-dimensional ferromagnetic Ising model respectively. We compare the MC results with those of a renormalization-group (RG) study where the critical fixed point is treated as a perturbation of the MF one, along the same lines as in the $\epsilon$-expansion for the Ising model. The MC and the RG method agree in the MF region, predicting the existence of a transition and compatible values of the critical exponents. Conversely, the two approaches markedly disagree in the NMF case, where the MC data indicates a transition, while the RG analysis predicts that no perturbative critical fixed point exists. Also, the MC estimate of the critical exponent $\nu$ in the NMF region is about twice as large as its classical value, even if the analog of the system dimension is within only $\sim 2\%$ from its upper-critical-dimension value. Taken together, these results indicate that the transition in the NMF region is governed by strong non-perturbative effects.
\end{abstract}

\maketitle
 
The renormalization group (RG) is a powerful method to tackle the complexity of a variety of thermodynamical systems close to their critical point: with the RG, an extensive number of microscopic variables is reduced to a few coarsened degrees of freedom whose behavior can be completely characterized in the thermodynamic limit. The RG theory for homogeneous systems---such as the $d$-dimensional Ising model on a hypercubic lattice---is based on a perturbative expansion around mean-field (MF) theory. Corrections to the MF fixed point (FP) are described in terms of a power series in $\epsilon = 4-d$ \cite{wilson1974renormalization}, and the predictions of this expansion in three dimensions have been shown to be in excellent agreement with both numerical simulations \cite{baillie1992monte} and experimental measurements from a wide variety of systems---binary fluids, superfluid helium, and  ferromagnets \cite{transitions1980vol}. For inhomogeneous, disordered systems characterized by a complex energy landscape such as Ising spin glasses \cite{edwards1975theory}, the formulation of a RG theory is  more involved: perturbative RG studies based on the $\epsilon$-expansion \cite{chen1977mean,kotliar1983one} are limited by the intricate diagrammatic structure of the underlying field theory, and the markedly divergent behavior of the $\epsilon$-expansion prevents any physical predictions below the upper critical dimension \cite{castellana2011renormalization}.

In the last few decades, perturbative RG approaches have  been extensively applied  to a disordered system of particular relevance in condensed-matter physics: an Ising spin-glass in the presence of an external magnetic field \cite{binder1986spin}. Indeed, the existence of a spin-glass transition in a field has been raising particular interest \cite{katzgraber2009study,leuzzi2009ising,houdayer1999ising,banos2012thermodynamic} because evidence in favor or against this transition would provide new insights into the structure of the low-temperature equilibrium states of the system \cite{fisher1986ordered}. The spin-glass transition in a field has been extensively studied with RG methods by examining a tentative set of critical FPs \cite{pimentel2002spin,bray1980renormalisation,moore2011disappearance}, although the existence of such a transition below the upper critical dimension is still under debate  \cite{moore2011disappearance,parisi2012replica}. In this regard, a recent RG study \cite{castellana2015hierarchical} has shown for the first time that the complete set of perturbative FPs in a field can be determined  for a spin glass built on a hierarchical lattice \cite{franz2009overlap}: besides the well-known trivial FP in the MF region---corresponding to $d\geq4$ for a ferromagnetic Ising model in $d$ dimensions---this analysis demonstrated by means of an $\epsilon$-expansion that there is no perturbative FP in the non-mean-field (NMF) region \cite{franz2009overlap}. The latter result provides a natural starting point for testing non-perturbative effects in spin glasses in a field: given that the RG analysis determined the complete set of perturbative FPs, if another FP in the NMF region  were found with a non-perturbative approach, then this FP would necessarily lie outside the perturbative regime.

In this paper, we investigate these non-perturbative effects by studying with Monte Carlo (MC) simulations the hierarchical Edwards-Anderson model in a magnetic field (HEA)---a pairwise spin-glass model with binary couplings and magnetic fields where spin couplings are disposed in a hierarchical way. The HEA and the hierarchical model in the RG study above \cite{castellana2015hierarchical} share the same distance dependence of the  coupling strength between spins: this common feature between the two models  allows us to compare the predictions of MC simulations with those from the  RG approach. Specifically, we study the existence of a phase transition by analyzing the temperature dependence of the correlation length and spin-glass susceptibility, and we characterize the value of the critical exponent $\nu$ \cite{wilson1974renormalization}. In the MF region, the MC data and the RG approach both predict that there is a transition in a field, and that this transition is characterized by a classical value of the critical exponent. Conversely, in the NMF region the two approaches exhibit a marked discrepancy: The RG method shows that there is no perturbative FP, while the MC data hints that there is a transition. In addition, the MC value of the critical exponent $\nu$ differs by nearly a factor of two from its classical RG value even if the analog of the system dimension is within only $\sim 2\%$ from its upper-critical-dimension value. Overall, these findings suggest that the phase transition in the NMF region is associated with a FP that cannot be considered as a perturbation of the MF FP, thus hinting at the presence of strong non-perturbative effects.

The rest of the paper is structured as follows: In Section \ref{results} we introduce the HEA, and in Section \ref{res1} we discuss how a suitable correlation length based on the hierarchical distance can be extracted from the correlation functions. In Section \ref{res2} we present the MC data for the correlation length and overlap probability distribution, and in Section \ref{res3} we discuss the numerical evidence for non-perturbative effects. Finally, in Section \ref{conc} we discuss the physical interpretation of our results and we lay out some topics of future studies. 

\section{Results}\label{results}

The  HEA is a system of $2^k$ Ising spins $S_i = \pm 1$, with Hamiltonian 
\be\label{bhea}
H_k[\vec{S}] = - \sum_{i<j=0}^{2^k-1} J_{ij} S_i S_j - \sum_{i=0}^{2^k-1} h_i S_i,
\ee
where $\{ J_{ij} \}$ are independent and identically distributed (IID) random variables defined as follows. Given two sites $i,j$, we denote by $d_{ij}$ the hierarchical distance between $i$ and $j$, i.e. the number of hierarchical levels that we need to ascend in the hierarchical tree starting from spins $i,j$ to find a root common to $i$ and $j$, see Fig. \ref{fig3}a. In Eq.  (\ref{bhea}), we then chose $J_{ij}$ to be different from zero  with probability 
\be\label{eq1}
p_{ij} = 2^{ -2 \sigma (d_{ij} - 1) },
\ee 
where nonzero $J_{ij}$s are equal to $\pm 1$ with equal probability, the parameter $\sigma$ determines how fast spin-spin interactions decrease with distance, and we chose $1/2 < \sigma < 1$  because this is the interval where the hierarchical Edwards-Anderson model is expected to have a finite-temperature phase transition \cite{franz2009overlap}. The magnetic fields $\{ h_i \}$ in Eq. (\ref{bhea}) are also IID random variables: $h_i$ is different from zero with probability $p_h$, and nonzero $h_i$s are equal to $\pm 1 $ with equal probability. 

The choice (\ref{eq1}) for the probability of placing a bond implies that the average interaction strength on hierarchical level $l>1$ between spins $S_i$ and $S_j$  is 
\be\label{eq50}
\mathbb{E}[J_{ij}^2] =  2^{-2 \sigma (l-1)},
\ee
where $\mathbb{E}[ \, ]$ denotes the average with respect to all random variables. Equation (\ref{eq50}) shows that, for a given $\sigma$, the power-law dependence on  $l$  of the average interaction strength is the same as in a HEA with Gaussian couplings studied recently \cite{castellana2015hierarchical}. The observation above suggests that the HEA introduced here and  the HEA with Gaussian couplings have the same critical features: this will be the working hypothesis of the rest of the paper. In particular, this assumption implies that the two models share the existence, or the absence, of a spin-glass transition, and that they both have a MF behavior characterized by vanishing order-parameter fluctuations for $1/2 < \sigma \leq 2/3$, and a NMF behavior in the region $2/3 < \sigma <1$ where order-parameter fluctuations arise \cite{franz2009overlap,castellana2011renormalization,castellana2010renormalization}.

We will now study the existence of a phase transition related to the divergence of the spin-glass susceptibility in the HEA: to do so, we will introduce the correlation length associated with the long-wavelength modes of the spin-glass susceptibility and we will study it as a function of temperature for different system sizes.

\subsection{Correlation length}\label{res1}

Since the probability (\ref{eq1}) with which non-zero bonds $J_{ij}$ are drawn in the Hamiltonian (\ref{bhea}) is a function of the hierarchical distance $d_{ij}$, the correlation function \cite{katzgraber2005probing}
\be\label{eq46}
\Gamma_{ij} \equiv \mathbb{E} \big[\left( \langle S_i S_j \rangle - \langle S_i \rangle  \langle S_j \rangle \right)^2  \big]
\ee
depends on $i$ and $j$  only through $d_{ij}$, where $\langle \, \rangle$ denotes the Boltzmann average with Hamiltonian (\ref{bhea}) and inverse temperature $\beta = 1/T$. To obtain the correlation length, we recall that the hierarchical distance $d_{ij}$ is related \cite{parisi2000p} to the dyadic norm $\left| \phantom{\;} \right|_2$ by the relation
\be\label{eq100}
\left| \mathcal{I}(i) - \mathcal{I}(j) \right|_2 = 2^{k-d_{ij}},
\ee
 where
\be\label{eq54}
\mathcal{I}(i) \equiv \sum_{j=0}^{k-1} a_{k-1-j} 2^j, \;\; i = 0, \ldots,   2^k -1, 
\ee 
and the coefficients $\{ a_j \}$ are given by the expression in base two of $i$: 
\be\label{eq55}
i = \sum_{j=0}^{k-1} a_{j} 2^j.
\ee
Equation (\ref{eq100}) implies that the correlation function $\Gamma_{ij}$ depends on $i$, $j$ only through $\left| \mathcal{I}(i) - \mathcal{I}(j) \right|_2$:
\be\label{eq45}
\Gamma_{ij} = f(\left| \mathcal{I}(i) - \mathcal{I}(j) \right|_2).
\ee

\begin{figure}
\centering\includegraphics[scale=0.5]{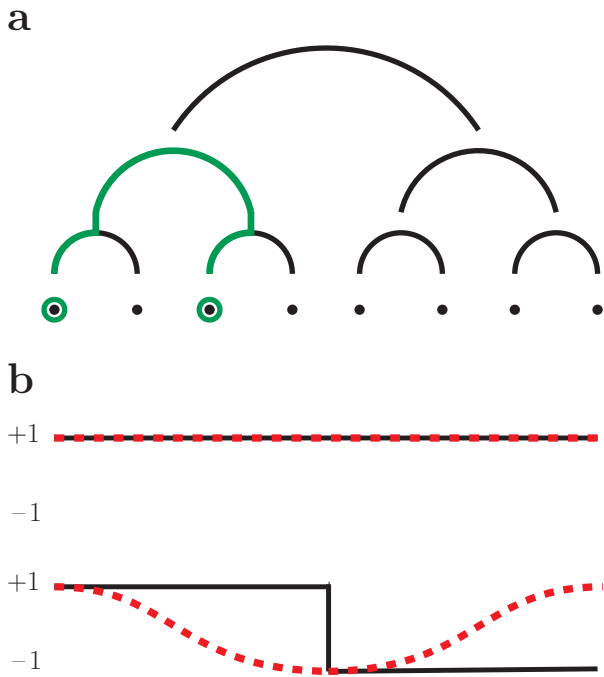}
\caption{Hierarchical Edwards-Anderson model in a magnetic field (HEA) with $k=3$. (a) Structure of spin-spin interactions: each dot represents a spin, and arcs represent  interactions between spins below them. The green path goes from the bottom to the top of the hierarchical tree until a common arc between the two circled  spins is found: since this requires ascending two hierarchical levels, the hierarchical distance between the two spins circled in green is $d_{ij} = 2$. (b) Long-wavelength Fourier modes: real part of the Fourier modes $\exp ( 2 \pi \imi p \,  \mathcal{I}(i) /2^k )$ for the HEA (black solid lines), where the index $i = 0, \ldots, 2^k-1$ runs over the lattice sites from left to right, and the Fourier modes take values $1$ and $-1$. We also plot the real part of the Fourier modes $\exp ( 2 \pi \imi p  \, i / 2^k )$ for a one-dimensional spin glass where spin-spin interactions decay according to the Euclidean distance $|i-j|$ (red dashed lines).   Top: first Fourier mode, corresponding to $p  = 0$ for both the HEA and for the power-law spin glass. Bottom: second Fourier mode, corresponding  to $p = 2^{k-1}$ for the HEA and to $p = 1$ for the power-law spin glass. \label{fig3}}
\end{figure}

We now relabel  the lattice indexes $i \rightarrow \mathcal{I}(i)$ and we perform the Fourier transform of $\Gamma_{ij}$ \cite{castellana2011renormalization}: this leads to the definition of the momentum-dependent spin-glass susceptibility
\bea\label{eq43}
\chi_{\rm SG}(p) & = & \frac{1}{2^k} \sum_{i,j=0}^{2^k-1} \exp\left( \frac{2 \pi \imi p }{2^k}  (i-j) \right) \Gamma_{\mathcal{I}(i) \mathcal{I}(j)} \\ \nn
& = & \frac{1}{2^k} \sum_{i,j=0}^{2^k-1} \exp\left( \frac{2 \pi \imi p}{2^k} (i-j) \right)  f(\left| i - j \right|_2),
\eea
where $p = 0, \ldots, 2^k-1$ is an integer momentum variable, and in the second line we used Eq. (\ref{eq45}) and the identity $\mathcal{I}(\mathcal{I}(i)) = i$. 

In spin-glass models with nearest-neighbor interactions, the correlation length is extracted \cite{ballesteros2000critical} from the Fourier transform of the correlation function by considering the values of $p$ with the smallest norm $|p|$. For the HEA, the natural definition of distance between sites on the lattice is not the Euclidean norm $|p|$, but the dyadic norm $|p|_2$ \cite{castellana2011renormalization}: to extract the long-wavelength behavior, it is thus natural to consider the  values of $p$ with the smallest value of $\left| p \right|_2$. 
To this end, we observe that the spin-glass susceptibility (\ref{eq43}) is the Fourier transform of a function $f$ which depends on $i-j$ only through its dyadic norm, thus  $\chi_{\rm SG}(p)$ depends on $p$ only through the dyadic norm $\left| p \right|_2$ of the momentum \cite{parisi2000p}. The long-wavelength behavior is then encoded into the limit $\left| p \right|_2 \rightarrow 0$, in which Eq. (\ref{eq43}) takes the modified Ornstein-Zernike form \cite{katzgraber2005probing}
\be\label{eq44}
\chi_{\rm SG}(p) \propto  \frac{1}{\xi_k^{-(2 \sigma-1)} + \left| p \right|_2^{2\sigma-1}}. 
\ee 

To extract $\xi_k$ from Eq. (\ref{eq44}), we consider the momentum values with the smallest dyadic norm: these are $p = 0$ and $p = 2^{k-1}$, and their dyadic norms read $\left| 0 \right|_2 = 0$ and $| 2^{k-1} |_2 = 2^{1-k}$ respectively \cite{parisi2000p}. The correlation length $\xi_k$ is then obtained by considering Eq. (\ref{eq44}) for these values of $p$:  
\be\label{xi}
\xi_k = 2^{k-1} \left( \frac{\chi_{\rm SG}(0)}{\chi_{\rm SG}(2^{k-1})} - 1 \right)^{\frac{1}{2\sigma-1}}. 
\ee

We will now discuss a  convenient way of computing  the correlation length in terms of  products of spin overlaps \cite{mezard1987spin}, and thus extract $\xi_k$ from the MC data. We introduce a set of replicas $\{ \vec{S}^a \}$ of the spin configuration $\vec{S}$, and we set 
\be\label{eq51}
Q_{ab}(p) \equiv \frac{1}{2^k} \sum_{i=0}^{2^k-1} S^a_i S^b_i \exp\left( \frac{2 \pi \imi p}{2^k} \mathcal{I}(i) \right) . 
\ee
By using Eqs. (\ref{eq43}), (\ref{eq51}), we have 
\bea\label{eq57}\nn
\chi_{\rm SG}(p)  &  = &   2^k \{ \mathbb{E}[\langle Q_{12}(p) Q^{\ast}_{12}(p) \rangle]   -2 \, \mathbb{E}[\langle Q_{12}(p) Q^{\ast}_{23}(p) \rangle] + \\ 
&& + \mathbb{E}[\langle Q_{12}(p) Q^{\ast}_{34}(p) \rangle] \}, 
\eea
where in what follows the average $\langle \,  \rangle$ of a function of $\vec{S}^1, \vec{S}^2, \ldots$ denotes the Boltzmann average over all replicas \cite{mezard1987spin}, each replica having an independent Boltzmann measure with Hamiltonian (\ref{bhea}) and inverse temperature $\beta$. We now introduce the overlaps in the left and right half of the lattice
\be\label{eq56}
Q^L_{ab} \equiv \frac{1}{2^{k-1}} \sum_{i=0}^{2^{k-1}-1} S^a_i S^b_i, \,\,\, Q^R_{ab} \equiv \frac{1}{2^{k-1}} \sum_{i=2^{k-1}}^{2^k-1} S^a_i S^b_i,
\ee
and by using Eqs. (\ref{eq51}), (\ref{eq56}), we have 
\be\label{eq58}
Q_{ab}(0) \equiv Q_{ab} =  \frac{1}{2^k} \sum_{i=0}^{2^k-1} S^a_i S^b_i = \frac{Q^L_{ab} + Q^R_{ab}}{2},
\ee 
\bea\label{eq52}
Q_{ab}(2^{k-1}) & = & \frac{1}{2^k} \sum_{i=0}^{2^k-1} S^a_i S^b_i \exp\left( \pi \imi \,  \mathcal{I}(i) \right)  \\ \nn
& = & \frac{1}{2^k} \sum_{i=0}^{2^k-1} S^a_i S^b_i (-1)^{  \mathcal{I}(i)}  \\ \nn
& = & \frac{1}{2^k} \left( \sum_{i=0}^{2^{k-1}-1} S^a_i S^b_i - \sum_{i=2^{k-1}}^{2^k-1} S^a_i S^b_i  \right) \\ \nn
& = & \frac{Q^L_{ab} - Q^R_{ab}}{2},
\eea
where in the third line we observed that $\mathcal{I}(i)$ is even if $0 \leq i \leq 2^{k-1}-1$ and  odd if $2^{k-1} \leq i \leq 2^k-1$, see Eqs. (\ref{eq54}), (\ref{eq55}). By using Eqs. (\ref{xi}), (\ref{eq57}), (\ref{eq58}), (\ref{eq52}) we obtain the final expression for the correlation length
\bw
\bea\label{eq59}
\xi_k = 2^{k-1} \left(\frac{\mathbb{E}[\langle (Q^L_{12}+Q^R_{12})^2 \rangle] - 2 \,  \mathbb{E}[\langle (Q^L_{12}+Q^R_{12})(Q^L_{23}+Q^R_{23}) \rangle] + \mathbb{E}[\langle (Q^L_{12}+Q^R_{12})(Q^L_{34}+Q^R_{34}) \rangle]}{\mathbb{E}[\langle (Q^L_{12}-Q^R_{12})^2 \rangle] - 2 \,  \mathbb{E}[\langle (Q^L_{12}-Q^R_{12})(Q^L_{23}-Q^R_{23}) \rangle] + \mathbb{E}[\langle (Q^L_{12}-Q^R_{12})(Q^L_{34}-Q^R_{34}) \rangle]} -1\right)^{\frac{1}{2 \sigma -1}}. 
\eea
\ew
The derivation above provides a numerically convenient way of extracting the correlation length: to compute $\xi_k$, we do not need the full set of spin-spin correlations $\{ \Gamma_{ij} \}$, but only the  products of the left and right overlaps  in Eq. (\ref{eq59}).

The first two long-wavelength Fourier modes which determine $\xi_k$, i.e.  $\exp ( 2 \pi \imi p \, \mathcal{I}(i) /2^k )$ for $p= 0, 2^{k-1}$, are depicted in Fig. \ref{fig3}b: in particular,  the second Fourier mode is $(-1)^{\mathcal{I}(i)}$, and it is equal to $1$ and $-1$ in the left and right half of the lattice respectively, see Eq. (\ref{eq52}). This particular form of the Fourier modes is due to the hierarchical structure of interactions in the model: for example, a one-dimensional spin glass where interactions depend on the Euclidean distance rather than on the hierarchical distance \cite{kotliar1983one} possesses a different second Fourier mode, given by $\exp ( 2 \pi \imi \, i /2^k )$.

\subsection{Numerical results for the correlation length and overlap probability distribution}\label{res2}

We will now study the existence of a phase transition by analyzing the behavior of the correlation length as a function of temperature: namely, we will assume that if there exists a critical point, then for $T$ close to the critical temperature $T_c$ and large enough $k$ the following finite-size-scaling relations hold \cite{leuzzi2009ising,katzgraber2009study}
\bea\label{eq10}
\xi_k/2^{k\nu/3} =  g_{\rm MF}(2^{k/3}(T-T_c)),\; \;\; \;&& 1/2 < \sigma \leq 2/3,\\ \label{eq10b}
\xi_k/2^k =  g_{\rm NMF}(2^{k/\nu}(T-T_c)),\; \; \; \; && 2/3 < \sigma < 1,
\eea
where the correlation-length critical exponent exponent $\nu$ is equal to $1/(2 \sigma-1)$ for $1/2 < \sigma \leq 2/3$ \cite{castellana2011renormalization} and $g_{\rm MF}$, $g_{\rm NMF}$ are two different scaling functions in the MF and NMF region respectively. 

We computed the correlation length (\ref{eq59}) with MC simulations performed with the parallel-tempering algorithm \cite{swendsen1986replica}. For zero magnetic field the Hamiltonian (\ref{bhea}) is symmetric under a global spin flip, thus the second and third term in both the numerator and denominator of Eq. (\ref{eq59}) vanish and $\xi_k$ can be computed by simulating only two replicas per temperature, while for nonzero field four  replicas per temperature are needed.  We combined parallel tempering with the asynchronous multispin-coding method, where $64$ disorder samples are simulated simultaneously by encoding the values of the couplings $J_{ij}$ into a $64$-bit integer \cite{palassini1999universal}. This method allowed for equilibrating an extensive number of samples 
$8 \times 10 ^3 \leq S \leq 16 \times 10^3$ and system sizes $2^6, \ldots, 2^{13}$.

\begin{figure*}
\centering\includegraphics[scale=1.03]{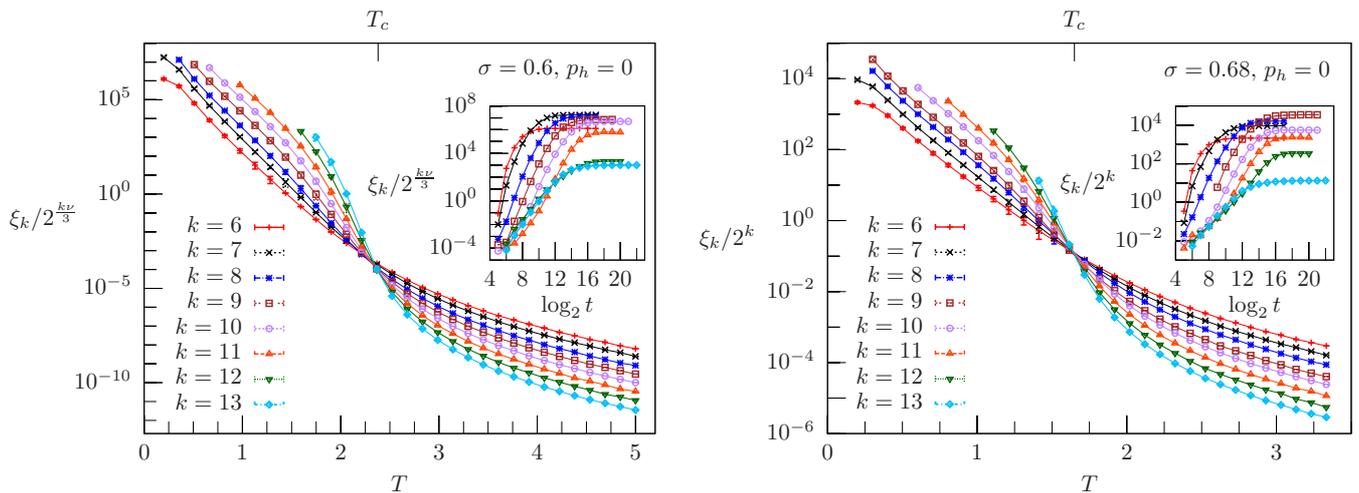}
\caption{
Correlation length for zero magnetic field, $p_h = 0$. Left panel: $\xi_k/2^{k \nu/3}$ vs. temperature $T$ in the mean-field region, $\sigma  = 0.6 < 2/3$, for system sizes $2^k$, with $k=6,\ldots, 13$ and $\nu = 1/(2\sigma-1)$: the curves cross at a critical temperature $T_c = 2.38 \pm 0.24$, which is also marked. Inset: $\xi_k/2^{k \nu/3}$ vs. $\log_2 t$ for the lowest temperature shown for each system size in the left panel, where $t$ is the number of Monte Carlo sweeps, and for every $t$ we plot $\xi_k/2^{k \nu/3}$ computed from the last $t/2$ sweeps. Right panel: $\xi_k/2^k$ vs. $T$  in the non-mean-field region, $\sigma  = 0.68 > 2/3$, with a critical-temperature value $T_c =  1.65 \pm 0.11$. Inset: $\xi_k/2^k$ vs. $\log_2 t$ for the lowest temperature shown for each system size in the right panel, where $\xi_k/2^k$ is computed as in the left-panel inset. \label{fig1}}
\end{figure*}

Let us first focus on the zero-field case $p_h=0$: in the left panel of Fig. \ref{fig1} we show $\xi_k/2^{k \nu/3}$ vs. $T$ in the MF region, $\sigma = 0.6$, while in the right panel we plot $\xi_k/2^k$ vs. $T$ in the NMF region, $\sigma = 0.68$. In the left and right insets we plot $\xi_k/2^{k \nu/3}$ and $\xi_k/2^k$  respectively vs. the logarithm of the number of MC sweeps, in order to provide an equilibration test for the data in the main panels.  The correlation-length curves cross at a finite critical temperature, showing that there is a spin-glass transition both in the MF and in the NMF case. This occurrence of a spin-glass transition in zero field is in agreement with a previous RG analysis for a HEA with Gaussian couplings and zero magnetic field, where a critical FP has been shown to exist both in the whole MF region $1/2 < \sigma \leq 2/3$ and in the NMF region $\sigma = 2/3 + \epsilon$ by means of an $\epsilon$-expansion \cite{castellana2011renormalization}.

\begin{figure*}
\centering\includegraphics[scale=1.25]{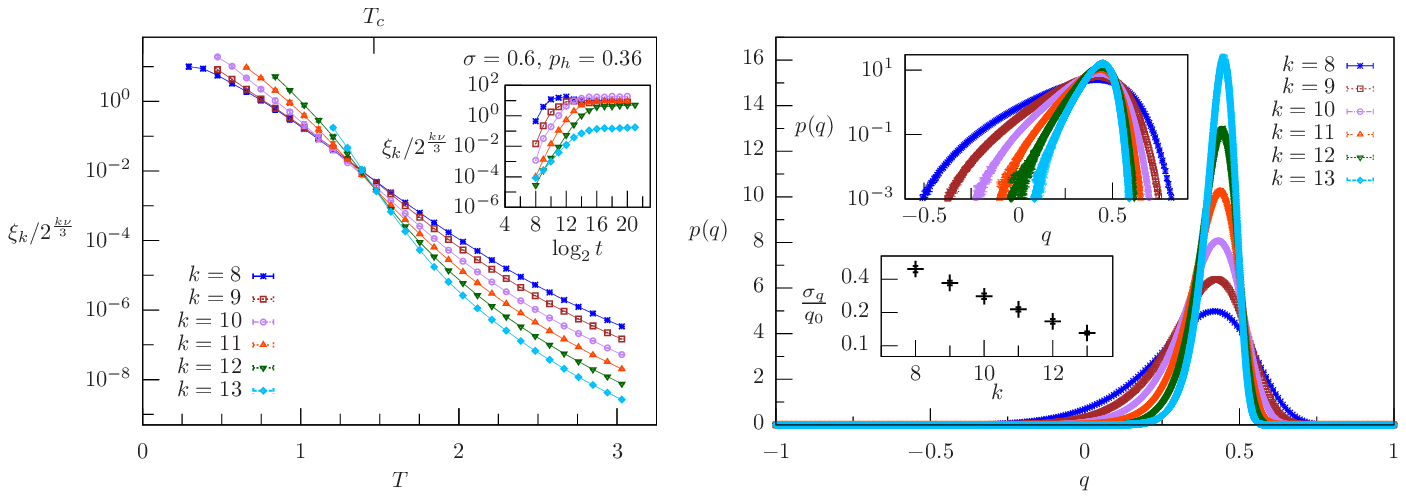}
\caption{
Correlation length and overlap probability distribution for nonzero magnetic field in the mean-field region, $\sigma  = 0.6 < 2/3$, with $p_h = 0.36$. Left panel: $\xi_k/2^{k \nu/3}$ vs. temperature $T$  and system sizes $2^k$, with $k = 8, \ldots, 13$.  Smaller sizes are affected by strong finite-size effects, and they are not shown in order to display clearly the curve crossings for larger sizes. The curves cross at a critical temperature $T_c = 1.46 \pm 0.07$, which is also marked. Inset: $\xi_k/2^{k \nu/3}$ vs. $\log_2 t$ for the lowest temperature shown for each system size in the left panel, where $t$ is the number of Monte Carlo sweeps, and for every $t$ we plot $\xi_k/2^{k \nu/3}$ computed from the last $t/2$ sweeps. Right panel: $p(q)$ vs. $q$ for $T = T_c$ and same system sizes as in the left panel. Top inset: same plot as in the right panel in logarithmic scale on the vertical axis. Bottom inset: ratio between the standard deviation $\sigma_q$ and the mean $q_0$ of $p(q)$ as a function of $k$. 
\label{fig2}}
\end{figure*}

\begin{figure*}
\centering\includegraphics[scale=3.]{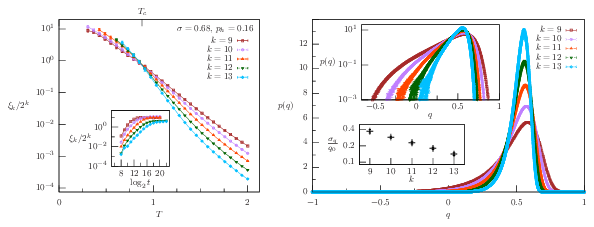}
\caption{Correlation length and overlap probability distribution for nonzero magnetic field in the non-mean-field region, $\sigma  = 0.68 > 2/3$, with $p_h = 0.16$. Left panel: $\xi_k/2^k$ vs. temperature $T$  and system sizes $2^k$, $k = 9, \ldots, 13$. Smaller sizes are affected by strong finite-size effects, and they are not shown in order to display clearly the curve crossings for larger sizes.  The curves cross at a critical temperature $T_c = 0.88 \pm 0.07$, which is also marked. Inset: $\xi_k/2^k$ vs. $\log_2 t$ for the lowest temperature shown for each system size in the left panel, where $t$ is the number of Monte Carlo sweeps, and for every $t$ we plot $\xi_k/2^k$ computed from the last $t/2$ sweeps. Right panel: $p(q)$ vs. $q$ for $T = T_c$ and same system sizes as in the left panel. Top inset: same plot as in the right panel in logarithmic scale on the vertical axis. Bottom inset: ratio between the standard deviation $\sigma_q$ and the mean $q_0$ of $p(q)$ as a function of $k$. 
\label{fig4}}
\end{figure*}

To study the nonzero-field case, we need to choose a value for the probability $p_h$ of placing a local magnetic field: it can be shown that a suitable value of $p_h$ is obtained as a tradeoff between two effects. First, for large $p_h$  the critical temperature---if any---is too low to achieve equilibration for sufficiently large system sizes. Second, for small $p_h$ a phase transition might appear, but this may be a spurious effect of the transition for $p_h=0$ shown in Fig. \ref{fig1} rather than an actual transition in a finite field. In this regard, a natural way of checking whether the apparent transition is an artifact of zero-field effects is to compute the overlap distribution
\be
p(q) \equiv \mathbb{E}\left[\langle \delta\left( q  - Q_{12} \right) \rangle \right].
\ee 
If there is a phase transition, in the thermodynamic limit $p(q)$ must converge to $\delta(q - q_0)$ at the critical point \cite{mezard1987spin}, where $q_0 \equiv \mathbb{E}[\langle Q_{12} \rangle]$, and  $q_0$ is positive because
\[
q_0 = \frac{1}{2^k} \sum_{i=0}^{2^k-1} \mathbb{E}[\langle S^1_i S^2_i \rangle]  = \frac{1}{2^k} \sum_{i=0}^{2^k-1} \mathbb{E}[\langle S_i  \rangle^2].
\]
When the system size is finite, the effect of the zero-field transition arises in the shape of a long exponential tail of $p(q)$ for $q<0$ \cite{katzgraber2009study,parisi2012numerical}: thus, a natural way of ensuring that we are in the true nonzero-field regime is to choose $p_h$ large enough that the support of $p(q)$ is mostly localized for positive $q$, i.e. $\sigma_q/q_0 \ll 1$, where $\sigma_q^2 \equiv \mathbb{E}[\langle (Q_{12} - q_0)^2\rangle]$.

We have thus chosen the value of $p_h$ according to the tradeoff above. In the MF region, $\sigma = 0.6$, we take $p_h = 0.36$: the left panel in Fig. \ref{fig2} shows that for this value of $p_h$ the curves of $\xi_k / 2^{k\nu/3}$ vs. $T$ cross at a finite critical temperature $T_c$. In addition, the right panel of Fig.  \ref{fig2} shows that for this value of $p_h$ the zero-field effects are weak: According to the top inset, the overlap distribution at $T_c$ has the typical left exponential tail due to zero-field effects \cite{parisi2012numerical}, nevertheless the main panel illustrates that the support of $p(q)$ is mostly localized at positive $q$, and the bottom inset shows that $\sigma_q / q_0 \ll 1$ for large $k$. The same qualitative scenario arises in the NMF region, $\sigma = 0.68$, see Fig. \ref{fig4}: The left panel  indicates that the curves $\xi_k / 2^k$ cross at a finite critical temperature, hinting at the existence of a phase transition in a field, and the right panel illustrates that such a transition is not an artifact due to the effects of the zero-field transition. 

\subsection{Hallmark of non-perturbative effects}\label{res3}

The correlation-curve crossings in the right panel of Fig. \ref{fig4} suggest that there is a spin-glass transition in the NMF region $\sigma > 2/3$: interestingly, this result is at variance with a recent perturbative RG analysis for a HEA with Gaussian interactions \cite{castellana2015hierarchical}, which shows with an $\epsilon$-expansion \cite{wilson1974renormalization} that for $\sigma = 2/3 + \epsilon$ there is no perturbative FP corresponding to a spin-glass transition in a field. The observation above thus raises the possibility that the phase transition in the right panel of Fig. \ref{fig4} results from non-perturbative effects, a scenario that we will investigate further in what follows.

\begin{figure*}
\centering\includegraphics[scale=1.025]{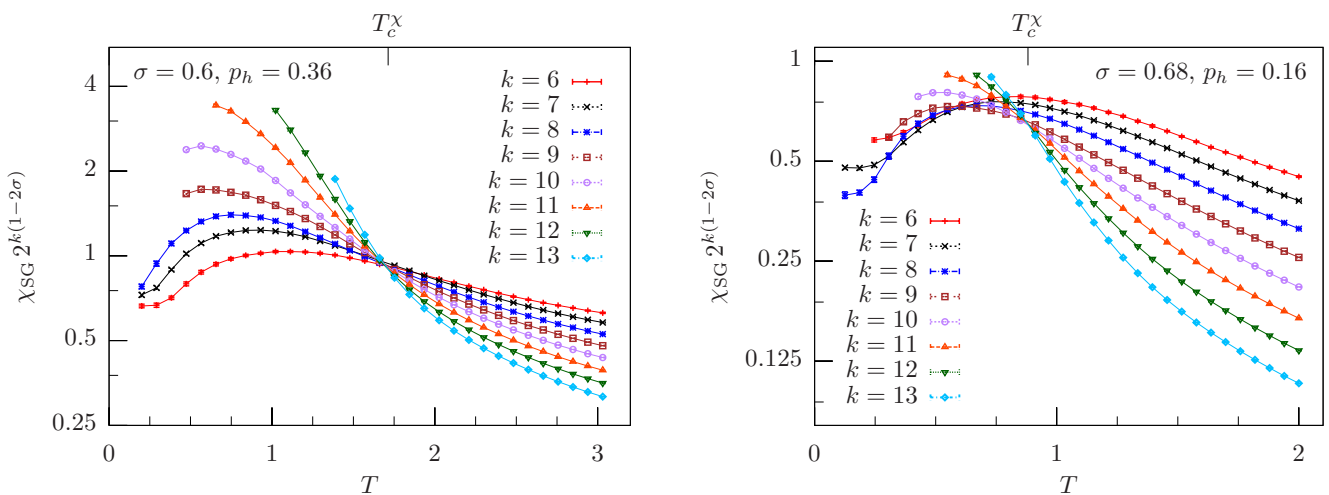}
\caption{Rescaled spin-glass susceptibility as a function of temperature for nonzero magnetic field. Left: $\chi_{\rm SG} \, 2^{k (1-2\sigma)}$ vs. $T$ in the mean-field region, $\sigma = 0.6 < 2/3$, for $p_h = 0.36$ and system sizes $2^k$, with $k=6, \ldots, 13$. Equilibration tests for $\chi_{\rm SG}$ have been performed along the same lines as  in Figs. \ref{fig1}, \ref{fig2}, \ref{fig4}, and they are not shown.  The curves cross at a finite temperature $T_c^\chi = 1.71 \pm 0.06$, which is also marked. Right: same plot as in the left panel in the non-mean-field region $\sigma = 0.68 > 2/3$ for $p_h = 0.16$ and  $k=6, \ldots, 13$. The crossing temperature of two curves with system sizes $2^k$ and $2^{k+1}$ is an increasing function of $k$, and the estimated infinite-volume critical temperature is  $T_c^\chi = 0.88 \pm 0.06$. 
\label{fig5}}
\end{figure*}

The  RG study mentioned above shows that if there is a spin-glass transition described by a perturbative FP, then at the critical point the rescaled spin-glass susceptibility $\chi_{\rm SG} \, 2^{k(1-2\sigma)}$ must converge to a finite value for $k \rightarrow \infty$, where $\chi_{\rm SG} \equiv \chi_{\rm SG}(0)$  \cite{castellana2015hierarchical}. The same RG analysis demonstrates that in the MF region there is a critical FP, hence in this region the rescaled susceptibility $\chi_{\rm SG} \, 2^{k(1-2\sigma)}$ must converge to a finite value at the critical temperature. This scenario is confirmed by the numerical data in the left panel of Fig. \ref{fig5},  which shows that in the MF region  $\chi_{\rm SG} \, 2^{k(1-2\sigma)}$ converges to a finite FP at a finite temperature $T_c^{\chi}$, which is within 
$\sim 17 \%$ from the critical temperature $T_c$ obtained from the correlation length in Fig. \ref{fig2}. The situation is completely different in the NMF region: there, the RG approach  shows that there exists no physical FP, thus suggesting that $\chi_{\rm SG}\,  2^{k(1-2\sigma)}$ does not  have a finite FP at any temperature \cite{castellana2015hierarchical}. Still, the right panel of Fig. \ref{fig5} illustrates that $\chi_{\rm SG} \, 2^{k(1-2\sigma)}$ converges to a finite FP at a finite temperature $T_c^{\chi}$. In addition,  $T_c^{\chi}$ coincides within error bars with the critical temperature $T_c$ obtained from the correlation length in Fig. \ref{fig4}. 

\begin{figure*}
\centering\includegraphics[scale=0.975]{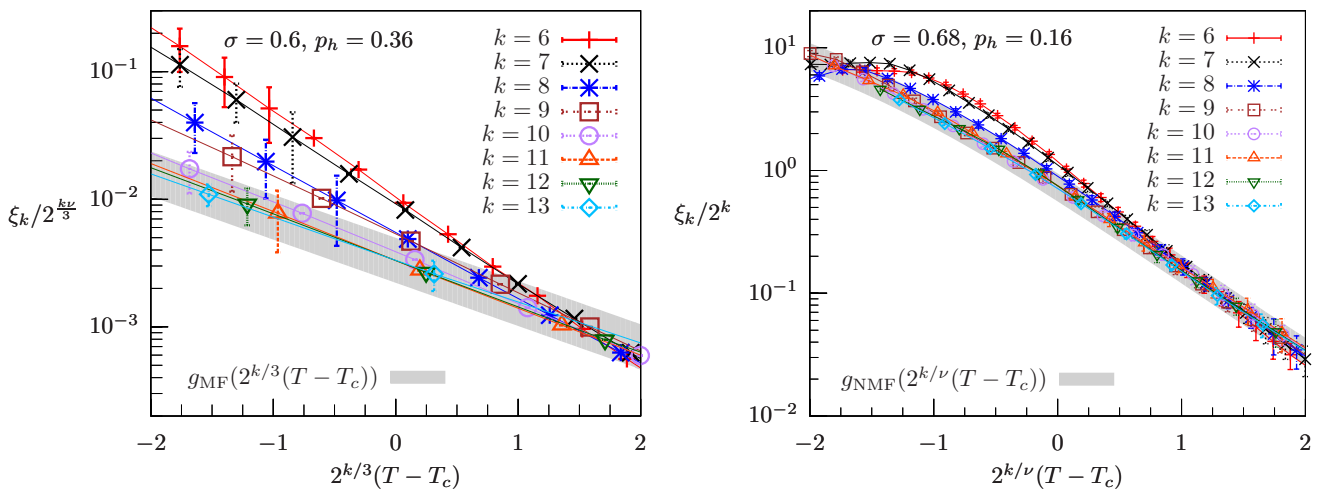}
\caption{Estimate of the critical exponent $\nu$ with a nonzero magnetic field. Left: $\xi_k / 2^{k \nu/3}$ vs. $2^{k/3}(T-T_c)$ in the mean-field region, $\sigma = 0.6 < 2/3$, for system sizes $2^k$, $k = 6, \ldots, 13$, where $\nu = 5.0$ and $T_c$ is determined by Fig. \ref{fig2}.  The gray line envelopes the curves for largest system sizes, and it provides an estimate of the finite-size-scaling function $g_{\rm MF}$. The value of $\nu$ has been chosen to achieve the best possible overlap between the curves enclosed in the gray envelope, yielding the estimate $\nu = 5.0 \pm 0.4$.  Right: $\xi_k / 2^k$ vs. $2^{k/\nu}(T-T_c)$ in the non-mean-field region, $\sigma = 0.68 > 2/3$, for system sizes $2^k$, $k = 6, \ldots, 13$, where $\nu = 5.0$ and $T_c$ has been determined by Fig. \ref{fig4}. The the finite-size-scaling function $g_{\rm NMF}$ is also shown, and $\nu$ has been estimated in the same way as in the left panel, providing  $\nu = 5.0 \pm 1.0 $.  
\label{fig6}}
\end{figure*}

Overall, the discrepancies between the predictions of the perturbative RG analysis and the numerical results in the NMF region raise the possibility that the phase transition resulting from the numerics in the NMF region is associated with a non-perturbative FP. In this regard, a general feature characterizing a critical FP is the exponent $\nu$  governing the exponential departure of RG trajectories as the temperature is moved away from its critical value \cite{wilson1974renormalization}. To study the existence of a non-perturbative FP, we estimate $\nu$ by means of the finite-size-scaling equations (\ref{eq10}), (\ref{eq10b}): we chose $\nu$ in such a way that for large $k$ the curves $\xi_k/2^{k \nu/3}$ vs. $2^{k/3}(T-T_c)$ and $\xi_k/2^k$ vs. $2^{k/\nu}(T-T_c)$ in the MF and NMF region respectively merge into a single curve.  In the MF region, $\sigma = 0.6$, we have $\nu = 5.0 \pm 0.4 $,  see Fig. \ref{fig6} left, in agreement with the RG classical value $\nu = 1/(2\sigma-1) = 5$ associated with the MF FP  \cite{castellana2011renormalization}. The situation is again very different in the NMF region: if we use the MF FP prediction $\nu = 1/(2\sigma-1)$ to estimate $\nu$ for $\sigma = 0.68$, we obtain $\nu \sim 2.78$. This values differs by nearly a factor of two from the MC estimate $\nu = 5.0 \pm 1.0 $ in the right panel of Fig. \ref{fig6},  even if $\sigma$ is within only $\sim 2\%$ from its upper-critical-dimension value $\sigma = 2/3$. This discrepancy suggests that the FP giving rise to the phase transition in the numerics cannot be regarded as a slight perturbation of a MF FP with classical exponents. Importantly, this last observation may be at the bottom of the overall disagreement between the numerics and the RG analysis: the advance here is that the RG approach may fail to capture the critical behavior in the numerics because it is based on the hypothesis that the FP governing the transition in the NMF region is a perturbation of the MF one. 

\section{Conclusions}\label{conc}

We investigated the existence of non-perturbative effects in spin glasses with a Monte Carlo (MC) study for an Ising spin glass in a field where pairwise couplings are disposed in a hierarchical way---the hierarchical Edwards-Anderson model in a magnetic field (HEA). The spin-interaction decay in the HEA is controlled by the parameter $\sigma$, which is reminiscent of the dimension $d$ for a ferromagnetic Ising model \cite{leuzzi1999critical}: in particular, the mean-field (MF)  and non-mean-field (NMF) regions $\sigma \leq 2/3$ and $\sigma > 2/3$ correspond to $d \geq 4$ and $d<4$ respectively \cite{franz2009overlap}. 
The HEA is  suitable for assessing the existence of non-perturbative effects for two reasons: First, the binary nature of couplings and magnetic fields makes the HEA  fit for non-perturbative numerical approaches, such as MC simulations  \cite{palassini1999universal}. Second, the hierarchical structure of the couplings reproduces the same average interaction structure as in a HEA with Gaussian couplings recently studied with perturbative renormalization-group (RG) methods \cite{castellana2015hierarchical}. This RG analysis provided the complete set of perturbative RG fixed points (FPs): in the MF region, it was shown that there is a FP corresponding to a phase transition, while no such FP exists in the NMF region. In the MF region the MC results are in agreement with the RG predictions: the numerics show the existence of a finite-temperature transition  where the correlation length and the spin-glass susceptibility become singular at large sizes, providing an estimate of the exponent $\nu$ in accordance with its classical value \cite{castellana2011renormalization}. Conversely, the MC and the RG approach markedly disagree in the NMF region, where the MC data for both the correlation length and spin-glass susceptibility support the existence of a finite-temperature transition. Also, the MC estimate of $\nu$ in the NMF region differs by nearly a factor of two from its classical value, even if $\sigma$ is within only $\sim 2 \%$  from  the value $\sigma = 2/3$  separating the MF and NMF regions. This picture is at variance with other systems where perturbation theory is well-behaved such as the $d$-dimensional Ising model, where the MC estimate of $\nu$ in $d=3$, i.e. within $25 \%$ from the upper critical dimension, differs by only $\sim 25 \%$ from its classical value \cite{baillie1992monte}. 
Taken together, these findings hint that the spin-glass transition resulting from the numerics in the NMF region cannot be captured with a perturbative framework, raising the possibility that this transition is related to a FP which cannot be described as a perturbation of the MF one.  

These findings raise a variety of possible scenarios concerning   the structure of the low-temperature phase. On the one hand, the occurrence of a transition in a field is associated with the replica-symmetry-breaking (RSB) scenario, a MF picture characterized by the existence of exponentially many low-lying energy states \cite{mezard1987spin}. Thus, the evidence provided here in favor of such a transition hints that the MF picture is accurate at least in some part of the NMF region. On the other hand, our findings show that the phase transition in the NMF region cannot be captured by an approach where corrections to the MF picture are considered as a perturbation. This fact raises the possibility that the physical features of this  transition are markedly different from the MF ones: this possibility is reminiscent of scenarios alternative to the RSB one, such as the droplet picture  \cite{fisher1986ordered} or other scenarios intermediate between the RSB and the droplet one \cite{krzakala2000spin}. In this regard, the structure of the low-lying energy states could be directly analyzed by probing the low-energy excitations above the ground state \cite{krzakala2000spin,hartmann2002large} and by characterizing, for example, their  stiffness \cite{fisher1988equilibrium} and surface dimension \cite{middleton2001energetics}. A interesting possibility would then be to study how the resulting picture for the low-energy landscape relates to the existence of non-perturbative effects.  Taken together, the points above provide an interesting direction that is worth exploring in future studies.

\begin{acknowledgments}
M. C. is grateful to M. A. Moore and E. Br\'{e}zin for useful comments and discussions. Research supported in part by NSF Grants PHY--0957573, PHY--1305525 and CCF--0939370, by the Human Frontiers Science Program, by the Swartz Foundation,  by the W. M. Keck Foundation, and by the European Research Council through grant agreement no. 247328--CriPherasy project. The simulations presented in this paper were performed on computational resources supported by the Princeton Institute for Computational Science and Engineering (PICSciE) and by the Office of Information Technology's High Performance Computing Center and Visualization Laboratory at Princeton University.
\end{acknowledgments}

\bibliographystyle{unsrt}
\bibliography{bibliography}

\end{document}